\documentclass[11pt]{article}
\usepackage{jheppub}
\usepackage{comment}

\newcommand{\al}{\alpha'}

\newtheorem{theorem}{Theorem}[section]

\newcommand{\be}{\begin{equation}}
\newcommand{\ee}{\end{equation}}

\newcommand{\boxit}[1]{%
  \[\fbox{%
      \addtolength{\linewidth}{-2\fboxsep}%
      \addtolength{\linewidth}{-2\fboxrule}%
      \begin{minipage}{\linewidth}%
      #1%
      \end{minipage}%
    } \nonumber \]%
}

\title{On the field theory expansion of superstring five point amplitudes}
\author{Rutger H. Boels}
\affiliation{II. Institut f\"ur Theoretische Physik, Universit\"at Hamburg\\ Luruper Chaussee 149, D- 22761 Hamburg, Germany }
\emailAdd{Rutger.Boels@desy.de}


\abstract{A simple recursive expansion algorithm for the integrals of tree level superstring five point amplitudes in a flat background is given which reduces the expansion to simple symbol(ic) manipulations. This approach can be used for instance to prove the expansion is maximally transcendental to all orders and to verify several conjectures made in recent literature to high order. Closed string amplitudes follow from these open string results by the KLT relations. To obtain insight into these results in particular the maximal R-symmetry violating amplitudes (MRV) in type IIB superstring theory are studied. The obtained expansion of the open string amplitudes reduces the analysis for MRV amplitudes to the classification of completely symmetric polynomials of the external legs, up to momentum conservation. Using Molien's theorem as a counting tool this problem is solved by constructing an explicit nine element basis for this class. This theorem may be of wider interest: as is illustrated at higher points it can be used to calculate dimensions of polynomials of external momenta invariant under any finite group for in principle any number of legs, up to momentum conservation. }

\begin{document}
\maketitle

\section{Introduction}

The ability to calculate is central to achieving understanding in theoretical high energy physics. For the specific case of scattering amplitudes, progress can be measured in general along the twin axes of numbers of loops and legs. In string theory, a third axis appears: that of orders in the field theory expansion. Whenever the momenta in a particular scattering experiment are small compared to the string scale, a perturbation theory may be set up. In practice, this is often referred to as the $\alpha'$ expansion. Furthermore, it is part of string theory textbooks to calculate scattering amplitudes in a flat background at the string tree level in the form of integrals over the boundary of the string worldsheet\footnote{In the closed (or mixed) case this follows after application of the Kawai-Lewellen-Tye \cite{Kawai:1985xq} relations (or their analogons \cite{Lancaster:1989qc}, \cite{Stieberger:2009hq})}. In principle the field theory expansion of the amplitudes follows from the integrals. Beyond the four point case however relatively little is known about these string scattering amplitudes and especially their field theory expansion. This is quite surprising, since these scattering amplitudes are one of the few things we know how to compute reliably in string theory and contains much information on the theory. This paper is part of a wider effort to address this unsatisfactory situation.

Considering the birthplace of string theory \cite{Veneziano:1968yb} it is humbling to think that the first systematic attempt at computing a five point superstring amplitude appeared only relatively recently in \cite{Medina:2002nk}. A first all multiplicity result for the first string theory correction at order $(\alpha')^2$ appeared in \cite{Stieberger:2006bh} in four dimensional kinematics. More simplifications for the integrand of the amplitudes were obtained in \cite{Mafra:2011nv}, where a particularly efficient way of ordering the open string theory amplitudes in terms of their field theory limits was introduced. Some first results for the structure of the field theory expansion appeared in e.g.  \cite{Barreiro:2005hv}, \cite{Stieberger:2009rr} and \cite{elvang}. A first systematic study of the field theory expansion beyond simple orders however was announced only very recently \cite{Schlotterer:2012ny}. The structure found there `experimentally' is currently only partially understood, see e.g. \cite{Drummond:2013vz}. One roadblock for further understanding is that the high order expansion results ($\sim \alpha'^{16}$ for five point amplitudes) of \cite{Schlotterer:2012ny} are not publicly available and are hard to obtain independently using current methods used in string theory. A prime goal of this article is to change this situation. 

It is known that for five point amplitude the functions which appear in the superstring are of the ${}_3 F_2$ hypergeometric type. Expansion of this type of function around integer arguments is a classical problem which appears often during the evaluation of complicated Feynman integrals in the more phenomenologically oriented sections of high energy physics. The literature on this is correspondingly vast: see for instance the references in \cite{Kalmykov:2007pf} and \cite{Bytev:2009kb}. The papers  \cite{Kalmykov:2007pf} and \cite{Bytev:2009kb} in particular introduce an expansion algorithm for hypergeometric functions which will be implemented in this article for superstring amplitudes with five massless legs. This algorithm highlights the close connection of the expansion to certain polylogarithms.  This is seen as a good starting point to deepen the investigation of the field theory expansion of the superstring. The techniques covered here do, in principle, generalize to higher leg multiplicity. Note that a naive field theory expansion does not make sense for amplitudes which involve any of the Regge excitations: their masses are given in string units in terms of $1/(\alpha')$.

This paper is structured as follows: In section $2$ some notation is established by writing several recently made conjectures. The expansion algorithm for open string amplitudes is subject of section $3$. In section $4$ the focus is on the closed string amplitudes which follow from the KLT relations. Special attention is given to a recently identified class of particularly simple amplitudes: the maximal R-symmetry violating ones. It is shown these take in the case at hand a particularly simple expression in terms of the open string amplitudes. To facilitate the discussion a simple nine element basis is introduced for the completely symmetric polynomials of external momenta. The used technology quite likely has applications beyond the immediate application at hand, as will be highlighted at the end of the section. A discussion and conclusion section round of the main presentation. The main results of this paper (explicit matrices, expansion coefficients, $\phi$-map) are contained in the archive submission for the readers convenience. The used algorithm is illustrated with an explicit example Mathematica implementation given in appendix A.

\begin{it} Note added in proof \end{it} \newline
While this paper was being readied for publication an unexpected overlap was discovered with a project by different authors which also aims at novel expansion methods for superstring amplitudes. Results from this project have since been announced in \cite{Broedel:2013tta} and \cite{them2}.

%%%%%%%%%%%%%%%%%%%%%%%%%%%%%%%%%%%%%%%%%%%%%%%%%%%%%%%%%%
%%%%%%%%%%%%%%%%%%%%%%%%%%%%%%%%%%%%%%%%%%%%%%%%%%%%%%%%%%

\section{Brief review of five point superstring amplitudes}

Many of the conjectures in this section originate in \cite{Schlotterer:2012ny}. Superstring amplitudes are functions of the external kinematics and polarizations. As shown in \cite{Mafra:2011nv} the polarization dependence of open superstring amplitudes resides completely in Yang-Mills tree amplitudes. The rest is a relatively simple function of the momenta of all particles $k^{\mu}_i$ subject to momentum conservation
\begin{equation}\label{eq:momconv}
\sum_{i=1}^n k^{\mu}_i = 0
\end{equation}
for $n$ incoming on-shell and massless ($\alpha'(k_i)^2= \frac{1}{4} s_{ii}=0$) momenta. Lorentz invariance then dictates that the basic variables are made of
\begin{equation}
s_{ij} = \alpha' (k_i + k_j)^2
\end{equation}
where $\alpha'$ is the usual dimension-full parameter which set the string scale. The dimensionless $s_{ij}$-variables will be referred to as Mandelstam variables or `Mandelstams' for any number of external legs. Momentum conservation implies relations between the $s_{ij}$. Let us first assume that the number of dimensions is larger than or equal to $n$. Then constraints on the Mandelstams can be derived from momentum conservation as displayed in equation \eqref{eq:momconv} by contracting with all $n$ possible momenta. Hence there are
\begin{equation}
\# \textrm{vars} = \frac{1}{2} n (n-3) 
\end{equation}
independent variables. For four points, these boil down to Mandelstam's original variables, subject to $s+t+u=0$. For five points an interesting set for our purposes is the cyclicly symmetric set
\begin{equation}
\{s_{12},s_{23},s_{34},s_{45},s_{51}\} \equiv \{s_1,s_2,s_3,s_4,s_5 \}
\end{equation}
The other $5$ Mandelstams can be expressed in terms of these by solving the momentum conserving equations as
\begin{equation}
\left\{\begin{array}{l} 
s_{13} =  -s_{12} - s_{23} + s_{45} \\ [2pt]
s_{14} =  -s_{51} +s_{23} - s_{45}  \\[2pt]
s_{24} =  s_{51} - s_{23} - s_{34}   \\[2pt]
s_{25} =  -s_{12} - s_{51} + s_{34}   \\[2pt]
s_{35} =  s_{12} - s_{34} -s_{45} 
\end{array}\right.
\end{equation}
For general multiplicity $n$ a set of independent variables can be chosen to be for instance
\begin{equation}\label{eq:solvmomconv}
\{ s_{ij}\, |\, i<j \leq n-2 \} \,\,\cup\,\, \{ s_{i, n-1}\, | \,1\leq i \leq n-3 \}
\end{equation}
which corresponds to the dependent variables
\begin{equation}\label{eq:solvmomconvII}
\{ s_{i,n}\, |\, i \leq n-1 \} \,\,\cup\,\, \{ s_{n-2, n-1} \}
\end{equation}
It is straightforward to check that the latter set of variables has a solution in terms of the former set under momentum conservation. Other sets are possible of course, but will not be needed here. There are more constraints when the number of legs is more than the number of dimensions plus one. This is simply due to the fact that $n$ vectors in a $D$ dimensional space are always dependent when $n>D$. Momentum conservation accounts for the `$+1$' offset. Since the subject of this note are superstring amplitudes, this will discrepancy will only kick in for $11$ or more particles and can safely be ignored for the present purposes.

\subsection{Open superstring}

The two independent five point open superstring amplitudes from which the others can be derived can be chosen to be the color-ordered amplitudes $A^{\textrm{o}}(12345)$ and $A^{\textrm{o}}(13245)$. These can be expressed in terms of the Yang-Mills amplitudes with the same ordering as
\begin{equation}
\left(\begin{array}{c} A^{\textrm{o}}(12345) \\ A^{\textrm{o}}(13245)  \end{array}\right) = F \left(\begin{array}{c} A^{\textrm{YM}}(12345) \\ A^{\textrm{YM}}(13245)  \end{array}\right)
\end{equation}
where
\begin{equation}\label{eq:Ffivepts}
F = \left(\begin{array}{cc} F_1 & F_2 \\ \tilde{F}_2 & \tilde{F}_1 \end{array} \right) 
\end{equation}
with the entries defined as the integrals
\begin{equation}\label{eq:F1}
F_1  =  s_{12} s_{34} \int_{0}^1\int_{0}^1 dx dy \,\,x^{s_{45}} y^{s_{12}-1} (1-x)^{s_{34}-1} (1-y)^{s_{23}} (1-xy)^{s_{24}}
\end{equation}
 and
 \begin{equation}\label{eq:F2}
 F_2  = s_{13} s_{24} \int_{0}^1\int_{0}^1 dx dy \,\,x^{s_{45}} y^{s_{12}} (1-x)^{s_{34}} (1-y)^{s_{23}} (1-xy)^{s_{24}-1}
 \end{equation}
and the functions $\tilde{F}_i$ are the $F_i$ with particles $2$ and $3$ interchanged.

\subsubsection*{MZVs and polylogarithms}
 The expansion of these functions in $F$  or indeed their higher point generalizations contains generically many so-called Multi-Zeta Values (MZVs). These are a class of numbers which appear in physics in several instances, either true classes of integrals or nested summations. They are specified by a vector of ``depth'' $d$ whose entries are positive integers that sum to the so-called ``weight''  $w$, i.e.
\begin{equation}
\zeta(n_1,n_2,\ldots {n_d}) 
\end{equation}
\begin{equation}
\sum_{j=1}^d n_j = w
\end{equation}
The weight of a product of MZVs is the sum of the weights. Weight will play a special role below as it will be shown to be correlated to order in the field theory expansion. In this context it is useful to note that weight is sometimes also referred to as transcedentality.  
  
Given a vector $\vec{n}$ one can construct a ordered vector $\vec{\tilde{n}}$ with entries $0$ or $1$ such that there are $n_i-1$ consecutive $0$'s follows by a one. 
\begin{equation}\label{eq:zerosones}
\vec{\tilde{n}} = \{0,0,\ldots, 1, 0, \ldots,1, \ldots \}
\end{equation}
Both bases will play a role in this article. 

The multi-zeta values are very well studied and obey many relations with values in the rationals $\mathbb{Q}$. An explicit basis under these relations up to and including weight $22$ exists for the MZVs \cite{datamine}: this basis will be referred to as the datamine basis.  Of central importance in this article is the realization that the multi-zeta values are boundary values of a class of functions called polylogarithms. To define these first specify an integral operator as
\begin{equation}
I(a): f(z) \rightarrow  \int_{0}^z dt \frac{1}{t-a} f(t)
\end{equation}
which takes a function $f(z)$ and integrates it to a new function $g(z)$. Polylogarithms $P$ are obtained by repeated application of this integral operator on $1$,
\begin{equation}
P(\vec{\tilde{n}},z) =  I(0)^{n_1-1} I(1) I(0)^{n_2-1} I(1) \ldots 1
\end{equation}
As the notation is intended to suggest  the values of the vector are  only $0$ or $1$, see equation \eqref{eq:zerosones}.  For the purposes of this article it is useful to define the MZVs as a boundary ($z=1$) value of these functions specified by a vector $\vec{\tilde{n}}$,
\begin{equation}
\zeta(\vec{\tilde{n}}) = \left[ I(0)^{n_1-1} I(1) I(0)^{n_2-1} I(1) \ldots 1 \right]_{z=1}
\end{equation} 
In more modern language, the vector $\vec{\tilde{n}}$ is essentially the so-called `symbol' of a particular harmonic polylogarithm. The appearance of MZVs in string theory is natural: it was noted long ago that the integrals appearing in string amplitudes can always be expressed as boundary values of functions \cite{mano}.

\subsubsection*{Definitions, conjectures}

The first conjecture is that the expansion of generic open superstring amplitudes at tree level is maximally transcendental: that is, the expansion at order $\alpha'^k$ have uniform transcedentality $k$. In other words, the conjecture is that the expansion at order $k$ is a function of momentum and polarization invariants and $\alpha'$ of the right mass dimension, multiplied by combinations of MZVs of total transcendentaility $k$ with rational coefficients. At order $2$ for instance, the only coefficient allowed to appear by this conjecture is $z(2) = \pi^2/6$. At higher orders similar results may be obtained: the non-rational coefficient appearing can always be expressed in terms of a basis of the MZVs. There is considerable support for the conjecture: it is known to be satisfied by four point tree level amplitudes for instance. Moreover, the results announced in  \cite{Schlotterer:2012ny} satisfy this. To our knowledge however, it has never been proven generically beyond four point amplitudes. 

Any known expansion algorithm for the integrals appearing in the five point amplitude will yield the answer in terms of generic MZVs which obey many relations. Using the datamine basis\footnote{Not forgetting that two MZVs with $\vec{\tilde{n}}_1$ and $\vec{\tilde{n}}_2$ are the same whenever one yields the other when interchanging ones and zeros and inverting the order. This is known as the 'duality' relation. } then gives an unambiguous answer which may be compared between approaches. In terms of this basis one can now define  two by two matrices $P_i, M_i$ as the coefficient of
\begin{equation}
P_{2i} = F |_{\zeta(2i) (\alpha')^{2i}}
\end{equation}
and 
\begin{equation}
M_{2i-1} = F |_{\zeta(2i-1) (\alpha')^{2i-1}}
\end{equation}
Conjecture: in terms of the $\alpha'$ expansion the function $F$ factorizes as
\boxit{\begin{equation}\label{eq:conj1}
F = B[P_{i}] C[M_i] \qquad \textbf{conjecture}
\end{equation}}
with
\begin{equation}
B = \left( \sum_{i=0}^{\infty} (\alpha')^{2i} \zeta_2^i P_{2i} \right) 
\end{equation}
Moreover, numerical coefficients $c_j$ exist such that
\boxit{\begin{equation}\label{eq:fconj}
C(M_i) = \sum_k (\alpha')^k \left(\sum_j c_j (M^{(k,j)})  \right) \qquad \textbf{conjecture}
\end{equation}}
where the sum ranges over the elements of the set  $M^{(j,k)}$. This set contains all products of $M_i$ matrices
\begin{equation}\label{eq:Mproduct}
M_{i_1} \cdot \ldots M_{i_j}
\end{equation}
such that the restriction 
\begin{equation}
\sum_j i_j = k
\end{equation}
holds, with the convention that $M_{\textrm{even}}$ and $M_1$ are zero. 

A second conjecture states that the coefficients $c_j$ trivialize under a map $\phi$ from the multi-zeta values to non-commutative polynomials constructed in \cite{brown}. These non-commutative polynomials are constructed out of non-commutative products of monomials $f_{2i+1}$ for $i>0$ and the commutative element $f_2 \leftrightarrow \zeta_2$. The conjecture is that for a product of any choice of $M$ matrices the coefficient is the noncommutative polynomial with the same indices as the chosen M's but the order inverted. Explicitly, the conjecture states that the coefficient $c_j$ of the product in equation \eqref{eq:Mproduct} obeys
\boxit{\begin{equation}\label{eq:conjphimap}
\phi(c_j) = f_{i_j} \star \ldots \star f_{i_1} \qquad \textbf{conjecture}
\end{equation}}
where $\star$ denotes the non-commutative product. 

The required $\phi$ map can be constructed recursively in weight, see \cite{brown} for details and results up to weight $10$ as well as \cite{Schlotterer:2012ny} for results to weight $16$. In practice a basis such as the datamine one is necessary to manage the needed algebra. Results for the $\phi$ map up to weight $21$ obtained by implementing the algorithm of \cite{brown} in Mathematica are included in the archive submission of this paper. Taken together these conjectures, if true, imply that once the matrices $M$ and $P$ are known the full superstring amplitude can be reconstructed.

\subsection{Closed superstring}
The KLT relations generically express closed superstring amplitudes in terms of open superstring amplitudes at the tree level. In the five point case this can be conveniently written as
\begin{equation}
A^{\textrm{cl}}_5 = \left(\begin{array}{c} A^{\textrm{o}}(12345) \\ A^{\textrm{o}}(13245)  \end{array}\right)^T \,\Sigma\,  \left(\begin{array}{c} A^{\textrm{o}}(12345) \\ A^{\textrm{o}}(13245)  \end{array}\right)
\end{equation}
where 
\begin{equation}
\Sigma = \left(\begin{array}{cc} \Sigma_{11} & \Sigma_{12} \\ \Sigma_{21} & \Sigma_{22} \end{array}\right) 
\end{equation}
with coefficients given in  \cite{Schlotterer:2012ny}. The field theory limit of this matrix will be denoted $\Sigma_0$. Two conjectures were made in \cite{Schlotterer:2012ny} about these matrices:
\boxit{\begin{equation}\label{eq:conj2}
P^T \Sigma  P = \Sigma_0 \qquad \textbf{conjecture}
\end{equation}}
and
\boxit{\begin{equation}\label{eq:conj3}
M_i^T \Sigma_0  = \Sigma_0 M_{i} \qquad \textbf{conjecture}
\end{equation}}
The conjecture in equation \eqref{eq:conj2} was verified up to and including order $17$, while the conjecture in equation \eqref{eq:conj3} was verified to and including order $16$. The first equation especially yields a significant simplification of closed string amplitudes.

%%%%%%%%%%%%%%%%%%%%%%%%%%%%%%%%%%%%%%%%%%%%%%%%%%%%%%%%%%
%%%%%%%%%%%%%%%%%%%%%%%%%%%%%%%%%%%%%%%%%%%%%%%%%%%%%%%%%%

\section{Field theory expansion of open string five point amplitudes}

In this section we show how the conjectures recalled above can be checked to high order. The driver here is the ability to expand the ${}_3F_2$ hypergeometric function which appear in the evaluation of the integrals in equations \eqref{eq:F1} and \eqref{eq:F2}. A neat way to obtain the hypergeometric function out of the integral is to use a Mellin-Barnes representation for the ``$(1-x y)^{\lambda}$'' type term in the integrand which reduces the integral to beta function type integrals. Performing this integral and re-writing the result as an infinite sum shows the coefficients of the integrals are such that the sum is a hypergeometric ${}_3F_2$ function. In particular, 
\begin{multline}\label{eq:F1hyp}
F_1 = \frac{\Gamma(1+s_1) \Gamma(1+s_2) \Gamma(1+s_3) \Gamma(1+s_4)}{ \Gamma(1+s_1 + s_2) \Gamma(1+s_3+s_4) }  {}_3 F_2 \left(\begin{array}{c}s_1,s_4+1, s_2+s_3 -s_5 \\1+ s_1+ s_2,1+ s_3+s_4 \end{array} ; z \right)\Huge{\lfloor}_{z=1}\end{multline}
\begin{multline}\label{eq:F2hyp}
F_2 = (s_4 -s_1-s_2)(s_5 -s_2-s_3)\frac{\Gamma(1+s_1) \Gamma(1+s_2) \Gamma(1+s_3) \Gamma(1+s_4)}{ \Gamma(2+s_1 + s_2) \Gamma(2+s_3+s_4) }\\  {}_3 F_2 \left(\begin{array}{c}1+s_1,1+s_4, 1+s_2+s_3 -s_5 \\ 2+s_1+ s_2, 2+s_3+s_4 \end{array}; z \right)\Huge{\lfloor}_{z=1}
\end{multline}
hold. Note that the expansion of the pre-factor in both cases is closely related to the four point amplitudes. These functions turn out to be a nice warmup example for expanding the ${}_3 F_2$ functions. 

The technique for expanding hypergeometric functions which appears in this section was first proposed in the field theory context in \cite{Kalmykov:2007pf} and \cite{Bytev:2009kb}. The reader is referred to these papers for a full explanation. In the course of this project also the HypExp 2.0 \cite{Huber:2007dx}, XSummer \cite{Moch:2005uc} and NestedSums \cite{WeinzierlNested} packages written in Mathematica, Form and C++ respectively were evaluated for the purpose of expanding the hypergeometric functions in equations \eqref{eq:F1hyp} and \eqref{eq:F2hyp}. For this particular application however these were slower by many orders of magnitude than the algorithm explained below. 

\subsection{Warmup: expanding the ${}_2F_1$ hypergeometric function}
Of interest here is the four point amplitude which follows from the ${}_2F_1$ hypergeometric function as:
\begin{equation}\label{eq:venezianoas2f1}
{}_2F_{1}(-a \al,b \al ;1+b\al;z)|_{z=1} = \frac{\Gamma(a\al+1) \Gamma(b\al+1)}{\Gamma(a\al+b\al+1)}
\end{equation}
where $\alpha'$ dependence has been emphasized. Hypergeometric functions generically obey differential equations as follows from their series expansion definition. In this case this equation is known as the hypergeometric differential equations and reads:
\begin{equation}
\left( z (\theta - \al \, a)(\theta + \al \, b) - \theta (\theta + \alpha' \, b) \right) {}_2F_{1}(-a \al,b \al ;1+b\al;z) =0
\end{equation}
where $\theta = z \frac{d}{dz}$. As above, we are eventually interested in the boundary value $z=1$. The trick is now to first expand in $\al$ before taking $\al \rightarrow 0$. Write for this expansion
\begin{equation}
{}_2F_{1}(-a \,\al,b \al ;1+b\, \al;z)  = \sum_{k=0}^{\infty} w_k(z) (\al)^k
\end{equation}
With this expansion the above differential equation can be written as a recursive differential equation,
\begin{equation}
(z-1) \theta^2 w_k(z) - \left(z (a-b) + b \right) \theta w_{k-1}(z) - z a b w_{k-2}(z) = 0 
\end{equation}
The first few terms can be determined directly from the expansion (or the integral): $w_0 = 1, w_1 = 0$. Furthermore, there are boundary conditions: 
\begin{equation}
w_k(0) = 0 \qquad (\theta w_k) (0) = 0 \qquad \forall k > 0
\end{equation}
With these boundary conditions the above equation can be solved by integrating the equation using the integrands suggested by the harmonic polylogarithms. In particular, note that the integral operator $I(0)$ is the inverse of the differential operator $\theta$
\begin{equation}
I(0) \odot \theta = 1
\end{equation}
up to boundary conditions. Repeated application of this equation reducesr the expansion of ${}_2F_{1}(-a \al,b \al ;1+b\al;z)$ to a 'symbol-level' recursion relation:
\begin{equation}\label{eq:rec2F1}
w_{j} = -a (\{0,1\} \cup (w_{j-1}) \backslash 1 ) + b (\{0\} \cup w_{j-1}) - ab (\{0,1\} \cup w_{j-2})
\end{equation}
Where the join operator is acting linear on the $\vec{\tilde{n}}$-type argument of the polylogs appearing in the previous coefficient and the $\backslash 1$ is a linear map which drop the leading entry of the $\vec{\tilde{n}}$ vector. In essence, this is the action of $\theta$ on an integral with first entry in the vector of $0$: it is easy to check that only terms of this type are generated.  The first few coefficients obtained by the recursion read:
\begin{equation}
\begin{array}{rl}
w_0  = &1\\
w_1  =& 0\\
w_2 = &- a b \, P({0,1}, z)\\
w_3  = & a b^2 \, P({0,0,1}, z) + a^2 b  \, P({0,1,1}, z) \\
w_4 =  &- a b^3 \, P({0,0,1}, z) - a^2 b^2 \, P({0,0,1,1}, z) - a^3 b \,  P({0,1,1,1}, z) 
\end{array} \end{equation}
where the reader is reminded that the coefficients $w$ depend on the parameter $z$. This parameter can be set to one after running the recursion relation, and specific MZVs are obtained.  A Mathematica implementation of this recursion relation is listed in appendix \ref{app:math}.

From this recursion relation one obtains immediately that the expansion of this function is maximally transcendental: the multi-zeta values which appear at order $k$ in the expansion have weight $k$ if those at weight $k-1$ do. Even in Mathematica the recursion relation is very fast: order $100$ needs about $1.5$ sec on an ordinary laptop. Furthermore, some experimentation yields the result that the expansion has the intriguing general form 
\begin{equation}
\frac{\Gamma(a\al+1) \Gamma(b\al+1)}{\Gamma(a\al+b\al+1)} = 1 - \sum_{i,j=1}^{\infty}  (- \alpha')^{i+j}\,  a^i \, b^j \, \zeta(\{0^i ,1^j\} )
\end{equation}
On other grounds, see e.g. \cite{Schlotterer:2012ny},  it is known that the expansion of the four point amplitude in the string theory can be expressed in terms of single zeta values only. Hence the above MZVs can be expressed in terms of single zetas only, a result which has indeed appeared in the math literature in \cite{broadhurst}.

\subsection{Expanding ${}_3F_2$ hypergeometric functions}
Now the same strategy can be employed for the ${}_3F_2$ hypergeometric functions in equation \eqref{eq:F1hyp} as well as in equation \eqref{eq:F2hyp}. Fundamental here is the expansion of the function
\begin{equation}\label{eq:fund3f2}
 {}_3 F_2 \left(\begin{array}{c}\alpha' a_1,\alpha' a_2,\alpha' a_3\\\alpha' b_1,\alpha' b_2 \end{array}; z \right) = \sum_k w_k(z) \alpha'^k
\end{equation}
The needed hypergeometric functions in the five-point amplitudes are related to this one by the relations \cite{Kalmykov:2007pf}
\begin{equation}
 {}_3 F_2 \left(\begin{array}{c}1+ \alpha' a_1,1+ \alpha' a_2,1+ \alpha' a_3\\ 2+ \alpha' b_1,2 +\alpha' b_2 \end{array}; z \right) = \frac{(1+ \alpha' b_1) (1+ \alpha' b_2) }{(\alpha')^3 a_1 a_2 a_3 \, z} \theta  {}_3 F_2 \left(\begin{array}{c}\alpha' a_1,\alpha' a_2,\alpha' a_3\\1+\alpha' b_1,1+\alpha' b_2 \end{array}; z \right) 
 \end{equation}
and
\begin{equation}
 {}_3 F_2 \left(\begin{array}{c}1+ \alpha' a_1, \alpha' a_2, \alpha' a_3\\ 1+ \alpha' b_1,1 +\alpha' b_2 \end{array}; z \right) = \frac{1}{\alpha' a_1} \left(\theta + \alpha' a_1\right)  {}_3 F_2 \left(\begin{array}{c}\alpha' a_1,\alpha' a_2,\alpha' a_3\\ 1+\alpha' b_1,1+\alpha' b_2 \end{array}; z \right) 
 \end{equation}
In passing we note that these relations show that the two by two matrix $F$ can be schematically be factorized as 
\begin{multline}
F = \left(\begin{array}{cc} \frac{\Gamma(1+s_2) \Gamma(1+s_3) \Gamma(1+s_4) \Gamma(1+s_5)}{ \Gamma(1+s_2 + s_3) \Gamma(1+s_4+s_5) }  & 0 \\ 0 & (2 \leftrightarrow 3)\end{array} \right) \\  \left( \left( \begin{array}{cc} {}_3F_2 & 0\\ 0 &   (2 \leftrightarrow 3) \end{array} \right)   +    \left( \begin{array}{cc} \frac{1}{\alpha' a_1^2 a_2 a_3} \theta {}_3F_2 &  \frac{1}{\alpha' a_1^2 a_2 a_3} \theta {}_3F_2\\  (2 \leftrightarrow 3)  &   (2 \leftrightarrow 3) \end{array} \right)   \right) \end{multline}
where the bottom line is the result of the exchange of particles $2$ and $3$ of the top line, with the line reversed. 

To write the analog of the recursion relation of equation \eqref{eq:rec2F1} above it is (also programming-wise) useful to introduce new variables for the function in equation \eqref{eq:fund3f2}
\begin{eqnarray}\nonumber
\Delta_1 = a_1 + a_2 + a_3 - b_1 - b_2\quad & \Delta_2 = a_1  a_2 + a_2 a_3+ a_3 a_1 - b_1 b_2 & \quad \Delta_3 = a_1  a_2  a_3 \\ &Q_1 = b_1 +b_2 \quad Q_2 = b_1 b_2 & \nonumber
\end{eqnarray}
which make obvious the usual symmetries of the function. Inverting the differential equations with boundary conditions then yields the recursion relation
\begin{multline}\label{eq:rec3F2}
w_{j} = \Delta_1 (\{0,0,1\} \cup (w_{j-1}) \backslash 2 )+   \Delta_2 (\{0,0,1\} \cup (w_{j-2}) \backslash 1 ) \\ + \Delta_3 (\{0,0,1\} \cup (w_{j-3})  - Q_1  (\{0\} \cup (w_{j-1}) -  Q_2  (\{0,0\} \cup (w_{j-2})
\end{multline}
with $w_0=1$ and $w_1 = w_2=0$ as boundary conditions. Recalling that the action of the $\theta$ operator simply removes a leading $0$ from the vector $\vec{\tilde{n}}$ of the polylog then yields a formula to expand the hypergeometric functions in equations \eqref{eq:F1hyp} and \eqref{eq:F2hyp}. Note that this makes it clear that the resulting expansion is maximally transcendental: if the expansion has this property to order $k$, then the above relation extends this to $k+1$. The base step in this inductive argument is trivial. A second feature of the above expansion is that since $w_3$ is proportional to $\Delta_3$, every coefficient in the expansion is. This can be used to verify that the expansion of the function in \eqref{eq:F1} and \eqref{eq:F2} is really a strict polynomial. Combining the recursion relation for the ${}_3 F_2$ hypergeometric function and its pre-factor is an interesting problem left to further research.

\subsection{Verifying conjectures up to and including weight $21$.}
Calculating the field theory expansion of the five point amplitude at a particular order now boils down to implementing the recursion relation and setting $z=1$ \emph{after} the last recursive step and \emph{after} taking any required $z$ derivatives. There is a FORM implementation of this algorithm \cite{kalmykovprivate} which has been used to get the expansion up to and including order $28$, in terms of series of terms of quite general MZVs. A less quick but still sufficient Mathematica implementation was used instead to verify the conjectures made by \cite{Schlotterer:2012ny}, recalled above, up to and including order $21$. It should be said that the bottleneck here is not expanding the hypergeometric functions but instead is the substitution of the datamine basis for the MZVs which creates memory issues\footnote{This restricts to roughly order $16$ on a machine with $8$GB of memory for a calculation which takes about $10$ minutes in Mathematica, much of which is used compiling MZV data.}.   

In general the landscape of results does not change from that sketched up to and including order $16$ \cite{Schlotterer:2012ny}. New is the appearance of relations amongst products of $M$ matrices, first at order $18$. These look remarkably like Jacobi-like relations for some algebra, but we have been unable to identify this algebra. At order $18$ the one relation appearing takes the form:
\begin{multline}
-M3\cdot M5\cdot M3\cdot M7 + M3\cdot M5\cdot M7\cdot M3 + M3\cdot M7\cdot M3\cdot M5 \\ - M3\cdot M7\cdot M5\cdot M3 + M5\cdot M3\cdot M3\cdot M7 - M5\cdot M3\cdot M7\cdot M3 \\- M7\cdot M3\cdot M3\cdot M5 + M7\cdot M3\cdot M5\cdot M3 = 0
\end{multline}
This relation was elegantly explained in \cite{Drummond:2013vz}. Similarly, there is one such relation at order $19$, three at order $20$ and six at order $21$. This lead to a further practical remark: in checking the conjecture of equation \eqref{eq:fconj} it turns out to be handy to first take the trace and then write the conjectured sum on the right hand side as a vector with the trace of each matrix a separate entry. Then this vector can be evaluated on the same number of data points as the length of the vector. This reduces inverting the problem to inverting a matrix, which is easily implemented in Mathematica\footnote{``LinearSolve'' is your friend here. Note this friend can solve equations involving non-invertible matrices (if the system has a solution). For speed, invert for every MZV appearing separately.}. Care has to be taken though to verify that not an accidentally degenerate set of data points was selected. Explicit results for the expansion coefficients as well as the $M$ and $P$ matrices are included in the source of the archive version of this article.

Based on the obtained expansion, the conjectures recalled in equations \eqref{eq:conj1},\eqref{eq:fconj},\eqref{eq:conj2} and \eqref{eq:conj3} can be verified up to and including weight $21$. The conjecture in \eqref{eq:conjphimap} however fails at order $21$. The failure appears for the coefficients of those matrix combinations which contain one $M_9$ and 4 $M_3$ matrices and the class with one $M_5$, one $M_7$ and three $M_3$ matrices: although simpler than the datamine basis expression, the polynomial expressions do not simplify to monomials. The explicit expressions can be reconstructed out of the supplied data in the archive version of this article, or by asking the author. 

Currently the author does not know why the discrepancy arises. Its origin could be in several places. First of all, this is the result of a complicated, computer-based calculation so human error should definitely not be ruled out: an independent check would be most welcome. It is most likely that an error would be in the transcription of the algorithm described in \cite{brown}. The expansion algorithm of the hypergeometric functions is fairly simple and no error is suspected for pushing from order $20$ to $21$: this was checked by comparing two different implementations of the algorithm. A human error could be found by independent calculation: the supplied data ($\phi$-map, amplitude expansion) allows a direct comparison. 

Apart from human error, there are could be subtleties in the exact procedure  used to regulate some of the integrals (with leading `ones') which appear in Brown's procedure\footnote{Technically, in the current implementation some symbols appearing in the map with `leading ones' which are divergent are reduced to leading zero type by using, in Brown's notation, $I(0;1;1)$=0, and the fact that the integrals satisfy a shuffle algebra. This regularization was suggested in \cite{Schlotterer:2012ny}}. Moreover, the implementation of Brown's algorithm is highly dependent on the MZV datamine basis: an error could have crept in here either in the published files or in the transcription of them into Mathematica-ready code.

%%%%%%%%%%%%%%%%%%%%%%%%%%%%%%%%%%%%%%%%%%%%%%%%%%%%%%%%%%
%%%%%%%%%%%%%%%%%%%%%%%%%%%%%%%%%%%%%%%%%%%%%%%%%%%%%%%%%%

\section{Application to closed string five point amplitudes in the MRV sector}

By the KLT relations, the previous results on field theory expanding the open superstring amplitudes have an immediate application in the field theory expansion of closed string amplitudes, of course also with five external points. Especially conjecture \eqref{eq:conj2} makes clear that there are large cancellations hidden in the KLT relations. Although explicit, the resulting expressions for the closed string amplitudes are generically complicated and not very illuminating without further insight. 

In this article the closed string amplitudes are therefore considered in a special sub-sector within the type IIB string in ten dimensions uncovered in \cite{Boels:2012zr} which is remarkably similar to MHV amplitudes in four dimensions. This class of amplitudes violates R-symmetry maximally, whereas the four dimensional MHV amplitudes violate helicity maximally. Structurally, these amplitudes can be expressed on an appropriate \cite{Boels:2012ie}  on-shell superspace as
\begin{equation}\label{eq:genericMRV}
A^{\textrm{MRV}} = \tilde{A} \, \delta^{16}(Q)
\end{equation}
where $\tilde{A}$ is a completely symmetric function of the external momenta which does not have massless poles for more than four particles. This implies in the field theory expansion that this function has no poles: it must be a strict polynomial.

By Bose symmetry MRV amplitudes in the field theory expansion are therefore expressible in terms of completely symmetric polynomia of the external momenta, as the fermionic delta function is completely symmetric. Lorentz invariance dictates that these must be polynomials in the Mandelstams, $s_{ij}$. The problem now is momentum conservation: solving this explicitly yields variables which generally do no transform nicely under permutations. This is already true in the four point case: one element of the set $s,t$ is, under permutations of the external legs certainly mapped to $u$. For four particles it is known that any completely symmetric polynomial $f$ can be written as a double polynomial expansion in two basis elements,
\begin{equation}\label{eq:basisfourpnts}
f(s,t,u) = \sum_{i,j=0}^\infty c_{ij} \sigma_2^i \sigma_3^j
\end{equation} 
for some coefficients $c_{ij}$ where
\begin{equation}
\sigma_2 = s^2 + t^2 + u^2 \qquad \sigma_3 = (stu)
\end{equation}
This will be re-derived below. 

\subsection{Classifying completely symmetric polynomials}

To the best of our knowledge a similar basis is unknown above four points in the existing literature. One can, in principle, construct completely symmetric polynomials by brute force by summing certain base polynomials over all permutations. Relations between these could be established by evaluating these, say, $k$ polynomials at $k$ different kinematic points by taking $s_{ij}$ to be natural numbers. The resulting $k$ by $k$ matrix has null eigenvectors which are the sought-for relations, as long as no roots of the polynomials are hit. Implementing this strategy for five points and comparing to the online integer sequence database \cite{Sloane:seq} then led to a far better approach through Molien's theorem. 

Molien's theorem might be of wider interest so is presented here in a general formulation:
\begin{theorem}
(Molien) Let $V$ be a vector space and $M$ a matrix representation of a group P, which is a subgroup of the permutation group on this space. Let $f_k$ be a function on the space $V$ invariant under the action of $M$, with fixed homogeneity e.g.
\begin{equation}
f_k(\lambda \vec{v}) = \lambda^k f_k(\vec{v})
\end{equation}
Let $d_k$ be the number of independent polynomials of degree $k$. Let $g(t)$ be the generating function for these numbers,
\begin{equation}
g(t) = \sum_k d_k t^k
\end{equation}
Then this generating function can be calculated as
\begin{equation}\label{eq:molien1}
g(t) = \frac{1}{|P|} \sum_{p \in P} \frac{1}{\det\left(\mathbb{I} - t M_{p}\right)}
\end{equation}
\end{theorem}
In the case at hand the subgroup of the permutation group is actually the permutation group of $n$ elements itself. 

The permutation group has a non-trivial action on the space of solutions to the momentum conservation constraints. This space can be spanned for instance by the set in equation \eqref{eq:solvmomconv}. To illustrate this in the four particle case, one can pick $s=(k_1 + k_2)^2$ and $t=(k_2 + k_3)^2$ as the independent variables. A permutation of particles one and two then induces
\begin{equation}
s \rightarrow s \qquad t \rightarrow -s-t
\end{equation}
which can be written in matrix notation as 
\begin{equation}
\left(\begin{array}{c} s\\ t \end{array}\right) \rightarrow \left(\begin{array}{cc} 1 & 0 \\ -1 & -1 \end{array}\right) \left(\begin{array}{c} s\\ t \end{array}\right) \equiv M_{1\leftrightarrow 2}  \left(\begin{array}{c} s\\ t \end{array}\right)
\end{equation}
It is straightforward to derive the action of all $24$ permutations in this case in the same way. This yields the Molien series for our problem for four particles as
\begin{equation}
g_4(t) = \frac{1}{24} \left( \frac{12}{1-t^2} + \frac{4}{1-2t + t^2} + \frac{8}{1+t+t^2} \right)
\end{equation}

Note that the terms in the Molien series in equation \eqref{eq:molien1} are constant on the different conjugacy classes. Hence more efficiently one can sum over only one representative $c$ of each conjugacy class $C(P)$ and multiply by the number of elements in this particular class,
\begin{equation}\label{eq:molien}
g(t) = \frac{1}{|P|} \sum_{c \in C(P)} \frac{\textrm{dim}(C_c(P)) }{\det\left(\mathbb{I} - t M_{c}\right)}
\end{equation}

The Molien series can be used to calculate the generating function for the number of completely symmetric, Lorentz invariant polynomials of the external momenta up to momentum conservation, ordered by the degree which in this case corresponds to twice the mass dimension. More generally, polynomia invariant under any subgroup of the permutation group can be obtained this way. Concretely, let the vector space $V$ be the set of independent Mandelstams \eqref{eq:solvmomconv} at a fixed multiplicity. The matrix representation of the permutation group can be computed for this set. Then equation \eqref{eq:molien} can be used to calculate the generating function an the sought-for numbers can then be read from its expansion. 

The explicit expressions for the generating functions get quite complicated quickly. The five particle result for instance is
\begin{multline}
g_5(t)  = \frac{1}{120}
   \left(\frac{24}{1-t^5}+\frac{30}{1-t^5-t^4+t}+\frac{20}{1-t^5-t^4-t^3+t^2+t} \right. \\ \left. +\frac{20}{1-t^5+t^4-t^3+t^2-t}+\frac{25}{1-t^5+t^4+2\, t^3-2\, t^2-t}\right. \\ \left. +\frac{1}{1-t^5+5 \,t^4-10 \,t^3+10 \,t^2-5\, t}\right)
\end{multline}
Of course, the above algorithm can easily be implemented in a symbolic computer program such as Mathematica. The results for the first $15$ degrees of the polynomials for up to and including $10$ particles are summarized in table \ref{tab:numbpolsfrommol}.
\begin{table}[!h]
\begin{tabular}{c|cccccccccccccccc}
\phantom{1} &  0 & 1 & 2 & 3 & 4 & 5 & 6 & 7 & 8 & 9 & 10 & 11 & 12 & 13 & 14  \\ \hline
4 &  1 & 0 & 1 & 1 & 1 & 1 & 2 & 1 & 2 & 2 & 2 & 2 & 3 & 2 & 3 \\
5 & 1 & 0 & 1 & 1 & 2 & 2 & 5 & 4 & 8 & 9 & 13 & 15 & 23 & 24 & 34  \\
6 & 1 & 0 & 1 & 2 & 4 & 6 & 13 & 19 & 36 & 58 & 97 & 149 & 244 & 364 & 558  \\
7 & 1 & 0 & 1 & 2 & 4 & 8 & 20 & 36 & 83 & 169 & 344 & 680 & 1342 & 2518 & 4695   \\
8 & 1 & 0 & 1 & 2 & 5 & 10 & 28 & 59 & 152 & 364 & 885 & 2093 & 4930 & 11199 & 25021  \\
9 & 1 & 0 & 1 & 2 & 5 & 10 & 31 & 72 & 205 & 557 & 1565 & 4321 & 11942 & 32131 & 84927  \\
10 & 1 & 0 & 1 & 2 & 5 & 11 & 33 & 81 & 246 & 722 & 2222 & 6875 & 21497 & 66299 & 202179 
\end{tabular}
\caption{Numbers of completely symmetric polynomials up to momentum conservation as a function of degree up to and including $10$ particles. \label{tab:numbpolsfrommol}}
\end{table}
Note that read vertically at column $i$ the series seems to stabilize at $2i-2$. This we have checked up to $16$ particles (disregarding Gramm determinant constraints). If this observation turns out to be true, this gives the numbers of polynomials up to degree $8$ as in table \ref{tab:stablenumbs}.
\begin{table}[!h]
\begin{tabular}{c|ccccccccc}
\phantom{1} &  0 & 1 & 2 & 3 & 4 & 5 & 6 & 7 & 8 \\ \hline
$n\geq16$ & 1 & 0 & 1 & 2& 5& 11&34&87 & 279
\end{tabular}
\caption{Conjectured asymptotic numbers of completely symmetric polynomials up to momentum conservation. \label{tab:stablenumbs}}
\end{table}

\subsubsection*{Basis}
In the four point case a basis for the completely symmetric polynomials is known as displayed in equation \eqref{eq:basisfourpnts}. The Molien series can be used to find a candidate for a similar basis at higher points. The argument proceeds degree by degree: first, at degree $2$ there is only one independent symmetric polynomial. Pick one representative of this. If a set of basis polynomials is known at degree $i$, then one can construct all polynomials at degree $i+1$ by taking products within the set such that the resulting degree is $i+1$. The difference between the number of polynomials thus generated and number obtained from the Molien series at this degree, if positive, is a lower bound on the number of polynomials which have to be added to the basis at degree $i+1$. It is a lower bound, as there can be relations between the products of basis-polynomials induced by momentum conservation. This algorithm can be used to estimate the number of different basis polynomials which are needed to generate all other polynomials. The results are listed in table \ref{tab:numberofbasispols} up to seven particles up to degree 18 polynomials.
\begin{table}[!h]
\begin{tabular}{c|ccccccccccccccccc}
 \phantom{1}& 2 & 3 & 4 & 5 & 6 & 7 & 8 & 9 & 10 & 11 & 12 & 13 & 14&15&16&17&18 \\ \hline
4&  1 & 1 & 0 & 0 & 0 & 0 & 0 & 0 & 0 & 0 & 0 & 0 & 0 & 0 & 0 & 0 & 0  \\
5& 1 & 1 & 1 & 1 & 2 & 1 & 1 & 1 & 0 & 0 & 0 & 0 & 0 & 0 & 0 & 0 & 0  \\
6& 1 & 2 & 3 & 4 & 7 & 7 & 12 & 11 & 16 & 4 & 11 & 0 & 0 & 0 & 0 & 0 & 0  \\
7& 1 & 2 & 3 & 6 & 14 & 22 & 48 & 85 & 163 & 247 & 469 & 497 & 692 & 0 & 0 & 0 & 0 
\end{tabular}
\caption{Numbers of needed basis polynomials as a function of degree based on numerology.\label{tab:numberofbasispols}}
\end{table}
For four and five particles we have checked up to degree $100$ polynomials that the indicated set generates enough polynomials at higher degree to explain the Molien series results. For four particles this just reconfirms the result expressed in equation \eqref{eq:basisfourpnts}. For five particles based on the presented numerology there is a candidate basis consisting of nine elements. These will be chosen to be
\begin{equation}
\tau_i = (\alpha')^i \left(s^i_{12} + \textrm{all permutations} \right) \qquad i\in \{2,3\ldots,9 \}
\end{equation}
as well as 
\begin{equation}
\tau_1 \equiv \tau_{6'} = (\alpha')^6 s^3_{12} s^3_{23}  + \textrm{all permutations} 
\end{equation}
We have checked explicitly to degree $9$ these nine elements indeed form a basis. Note that this does not exclude relations between polynomials generated by this set of nine basis elements at higher degree: the first instance of this is at degree $12$ where there is one relation between the $24$ possible polynomials. 

In conclusion, in this subsection it was argued that for five particles all symmetric polynomials $g$of the Mandelstams can be expressed in the $\tau$ basis just presented, i.e. there exist coefficients $c$ such that
\begin{equation}
g = \sum_{i_{1} \ldots i_{9}} c_{i_1,i_2,\ldots, i_9}\prod_{k=1}^{9} (\tau_k)^{i_k}
\end{equation}
If the polynomial $g$ is homogeneous of degree $k$ then
\begin{equation}
6 \, i_1 + \sum_{j=2}^9 j (i_j) = k
\end{equation}
should hold\footnote{Note this type of equation is neatly solved in Mathematica by the command ``FrobeniusSolve''.} for the integer valued coefficients $i_j$. This expansion is not necessarily unique from degree $13$ onwards.

\subsection{Obtaining explicit results for MRV amplitudes at five points}
The stage is now set to compute the MRV amplitudes at five points from the KLT relations and then express the result in the basis just constructed. This calculation was alluded to in \cite{Boels:2012zr} and will be spelled out below.

A generic consequence of the superspace form of the amplitude in equation \eqref{eq:genericMRV} is that the pre-factor $\tilde{A}$ multiplying the fermionic delta function follows basically from computing one component amplitude. In this case it is easiest to pick the amplitude with four gravitons and one holomorphic scalar out of the IIB spectrum\footnote{Calculation at low orders for these and similar amplitudes also appear in \cite{elvang}.}. The scalar is generically a combination of axion and dilaton, but the axion part is necessarily zero here (it would arise through KLT from open string amplitudes with a single fermion). The needed fermionic integral can be computed easily in a four dimensional approach that can be used here since the problem only involves five-particle kinematics. The result of this integral is, in spinor helicity notation, $\propto \langle 1 2\rangle^4 [ 34]^4 $ after assigning helicity $++$ to legs one and two and $--$ to legs three and four. This gives
\begin{equation}
\tilde{A} \langle 1 2\rangle^4 [ 34]^4 = A(++, ++, --,--,\phi)
\end{equation}
The right hand side of this equation may now be calculated by the KLT relations. The scalar $\phi$ decomposes in a sum over opposite gluon polarization for the fifth leg. Schematically, this reads
\begin{multline}
A_{\textrm{cl}}(++, ++, --,--,\phi) \sim \\ A_{\textrm{o}}(+, +, -,-,-)A_{\textrm{o}}(+, +, -,-,+) + A_{\textrm{o}}(+, +, -,-,+)A_{\textrm{o}}(+, +, -,-,-)
\end{multline}
where the open string amplitudes on the right hand side are color ordered. From the explicit expressions for the five point MHV amplitude and its conjugate it is easy to factor out the universal $\langle 1 2\rangle^4 [ 34]^4 $ factor. The resulting expression for $\tilde{A}$ is after some rewriting into Mandelstam invariants:
\begin{equation}\label{eq:explicitequation5pt}
\tilde{A} = \frac{1}{s_{12}s_{23}s_{34}s_{45}s_{51}}\left(G_{11} + \frac{s_{12}s_{34}}{s_{13} s_{24}} G_{22} - \frac{s_{12} s_{34} - s_{23} s_{14}+s_{13} s_{24}}{s_{13} s_{24}} G_{12} \right)
\end{equation}
where $G_{ij}$ refers to an entry from the matrix 
\begin{equation}
G = F^T \Sigma F = C^T \Sigma_0 C
\end{equation}
From this expression it is not obvious this is going to yield a completely symmetric \emph{polynomial} in the external momenta. This can be explicitly shown in examples though and is guaranteed by the underlying Bose symmetry of the on-shell superfield. From the obtained expansion of the open string five point amplitude it is now a simple matter to compute the MRV amplitude up to order $21$. Explicit expressions can readily be obtained by combination of equation \eqref{eq:explicitequation5pt} with the explicit results included in the archive version of this article. 

To write the algebraically complex result in terms of the basis of completely symmetric polynomials, one can employ the same numerical trick as mentioned before for verifying \eqref{eq:fconj}. This produces the coefficients of the field theory expansion of the MRV amplitudes into the $9$ element basis. Again, this is easily automated and explicit results can be obtained up to and including order $21$. For the readers convenience, the result up to and including order $\alpha'^{10}$ reads
\begin{multline}
\tilde{A} =  6 \,z(3) + \frac{5}{24} \tau_2 \, z(5)  - \frac{1}{12}\tau_3 \, z(3)^2 + \frac{7}{4608} \tau_2^2 \, z(7)+ \frac{7}{32} \tau_4 \, z(7)
 - \frac{5}{1728} \tau_2 \tau_3 \, z(3) \,z(5)\\  -  \frac{1}{20} \tau_5 \, z(3)\, z(5)  - \frac{1}{62208}\tau_2 \tau_4\, (90\, z(3)^3 - 793\, z(9)) +  \frac{1}{2985984 } \tau_2^3 \,(36 z(3)^3 - 247 \,z(9)) \\ + \frac{1}{7776} \tau_3^2 \,(9\, z(3)^3 - 14\, z(9))+ \frac{1}{2985984 }\tau_6\, (41472 \,z(3)^3 + 119808 \, z(9))  \\ + \frac{1}{2985984}\tau_{6'} \, ( 774144\, z(9) - 248832 \,z(3)^3 )-\frac{1}{576} \tau_2 \tau_5 \, z(5)^2 -  \frac{7}{331776 } \tau_2^2 \tau_3 \,z(3) \,z(7)  \\  - \frac{7}{2304} \tau_3 \tau_4 \,z(3) \,z(7)  - \frac{1}{28} \tau_7 \,z(3)\, z(7) + \mathcal{O} \left(( \alpha')^{11}\right)
\end{multline}
Higher orders get progressively more complicated and the author has been unable to establish a profound pattern. Brown's $\phi$ map also doesn't reveal any obviously nice structure, although the resulting polynomial is somewhat simpler. Noticeable is the generic appearance of MZVs with depth greater than 2, starting at order $(\alpha')^{11}$. It would be interesting to see if there is more structure to be found exploring the notion of bases of completely symmetric polynomials. For instance the basis employed above was chosen for constructional convenience only: it would be interesting to see if string theory would prefer one particular basis. Based on results so far no basis has been identified yet.

Based on the conjecture of \cite{Schlotterer:2012ny} one can also guess specific terms at, in principle, arbitrarily high order by plugging in appropriate powers of products of $M$ matrices into equation \eqref{eq:explicitequation5pt}. In particular coefficients of powers of single zeta's are readily isolated. As an example, consider the simplest series,
\begin{align}
\tilde{A}_{\zeta_3} & =  6 \\
\tilde{A}_{\zeta_3^2} & =  - \frac{1}{12}\tau_3 \\
\tilde{A}_{\zeta_3^3} & = \frac{1}{82944} \tau_{2}^3 +\frac{1}{864} \tau_{3}^2 - \frac{5}{3456 } \tau_{2} \tau_{4} + \frac{1}{72}\tau_{6} - \frac{1}{12} \tau_{6'} \\
\tilde{A}_{\zeta_3^4} & = - \frac{1}{5971968} \tau_{2}^3 \tau_{3} - \frac{1}{93312} \tau_{3}^3+ \frac{5}{248832 } \tau_{2} \tau_{3} \tau_{4}-  \frac{1}{5184} \tau_{3} \tau_{6} + \frac{1}{864} \tau_{3} \tau_{6'}
\end{align}
which shows only a hint of an underlying structure:
\begin{equation}
\tilde{A}_{\zeta_3^4} + \frac{\tau_3}{84} \tilde{A}_{\zeta_3^3} = \frac{1}{186624} \tau_3^3 
\end{equation}
We have no intrinsic understanding of this phenomenon. More general explicit results are easily established using the explicit results included in the archive version of this article.

%%%%%%%%%%%%%%%%%%%%%%%%%%%%%%%%%%%%%%%%%%%%%%%%%%%%%%%%%%
%%%%%%%%%%%%%%%%%%%%%%%%%%%%%%%%%%%%%%%%%%%%%%%%%%%%%%%%%%

\section{Discussion and conclusion}

In this paper the field theory expansion of superstring theory amplitudes with massless legs in a flat background has been discussed. The main idea which may generalize to higher points is to introduce an additional parameter into the integral (the `z' of the ${}_3 F_2$ function) and consider the field theory expansion for generic values of this parameters first. This idea certainly generalizes beyond the five point case and it will be interesting to see what further consequences may be derived from this. Using the Mellin-Barnes argument given at the start of section $3$ it is easy to see that above five points more complicated functions than hypergeometric ones will arise. These are generically boundary values of hypergeometric-like functions as discussed in \cite{mano}. The last cited paper provides a concrete platform to start exploring similar differential equations based approaches to expanding these functions to see what additional structure may arise. However, this will lead to far beyond the scope of this paper. 

Resolution of the discrepancy at weight $21$ between conjecture and calculated $\phi$ map is certainly called for. Furthermore, structurally it will be interesting to explore further applications of the recursive structure for the field theory expansion of the five point amplitudes. Here it would be most welcome to relate the appearing $\theta$ operator for instance to a natural world-sheet quantity. More generally, the structure of the field theory expansion encodes deep properties of the string theory and it is certainly worthwhile to explore these further, for instance starting with \cite{Barreiro:2005hv}. One direction here is reconstructing the effective action from the obtained field theory expansion. 

For MRV amplitudes an extension of arguments used in \cite{elvang} in the four dimensional case to IIB in $10$ dimensions would be most welcome to push beyond five legs. This is interesting for various reasons, not least because of the known connections between MZVs and modular functions \cite{Green:1997tv} which is certain to contain wonderful results waiting to be explored. It is expected that the multi-zeta value structure follows most natural in the context of MRV amplitudes in the closed IIB string. 

Molien's theorem certainly deserves to be wider known than it is currently in the physics community, especially in places where the permutation group makes an appearance. From the discussion above it should be obvious how to count and analyze much more general problems. Finally, it is hoped that the explicit results obtained in this article and included in the archive submission can be useful for others for applications beyond the ones mentioned here. 

\acknowledgments
It is a pleasure to thank Mikhail Kalmykov for collaboration in an early stage of this work and for sharing his insight into expansions of hypergeometric functions. Furthermore, I would like to thank James Drummond for discussions and Jos Vermaseren and David Broadhurst for correspondence. Oliver Schlotterer and a referee are thanked for comments on earlier versions of this article. This work was supported by the German Science Foundation (DFG) within the Collaborative Research Center 676 "Particles, Strings and the Early Universe". 

\appendix

\section{Mathematica implantation of the algorithm}\label{app:math}
In this appendix a Mathematica implementation is given of the expansion of the function 
\begin{equation}
{}_2F_{1}(e1 \al,e2 \al ;1+e2\al;z)|_{z=1} =  \sum_{k=0}^{\infty} w_k(1) (\al)^k
\end{equation}
as obtained in equation \eqref{eq:rec2F1}. The coefficients in this expansion are obtained as a list by the following lines of code:

\begin{verbatim}
Rem1[x_] := MZV[Drop[ToExpression[x], 1]]
Add[x_, y_] := y /. MZV[args_] -> MZV[Join[x, args]]

xp2F1 = Function[{e1, e2, order}, 
   Block[{ws = Table[0, {jk, 1, order + 1}]}, 
   ws[[1]] = 1 MZV[{}] ;  ws[[2]] = 0 ; 
    For[j =  3, j <= order + 1, j++, 
     ws[[j]] =  e1 Add[{0, 1}, (ws[[j - 1]] /. MZV -> Rem1)] 
     -  e2 Add[{0}, ws[[j - 1]]] + e1 e2 Add[{0, 1}, ws[[j - 2]]] // Expand]; 
     ws[[1]] = 1; ws]];
\end{verbatim}

\noindent The function appropriate for the four point amplitude, see equation \eqref{eq:venezianoas2f1}, then follows as
\begin{verbatim}
xp2F1[-s,t, order]
\end{verbatim}
where order is the order one requires. The output is a list of the coefficients $w_k$. Note the marked difference in timing with and without the `Expand' statement: this is a consequence of cancellations within the expansion which mathematica only takes into account by removing brackets. The expansion of the ${}_3F_2$ hypergeometric function given in equation \eqref{eq:rec3F2} can be implemented by suitably extended but very similar lines of code.

\bibliographystyle{JHEP}

\bibliography{5pt}

\end{document}